\documentclass[a4paper,12pt, epsfig]{article}
\def\letter{0}\def\pr{0}
\pdfoutput=1 
\usepackage{epsfig}
\usepackage{epstopdf}
\usepackage{graphicx}
\usepackage{ifthen}

\usepackage{amsmath}
\usepackage[psamsfonts]{amssymb}
\usepackage{euscript}

\usepackage{latexsym}
\usepackage[arrow,matrix,curve]{xy}

\newskip\humongous \humongous=0pt plus 1000pt minus 1000pt

\newif\ifdtup

\def\,{\hspace{-.1cm}}
\def\hsp{,\hspace{.7cm}}

\def\tf {\tilde{f}}
\def\ff{{\mathcal{F}}}

\def\fc#1#2 {\frac{n}{q}#1\frac{n}{q}#2}

\def\hf{H\p_{f_0}}

\newcommand{\vac}{\ensuremath{|0\rangle}}
\newcommand{\stt}{\ensuremath{|\psi\rangle}}

\renewcommand{\cos}{\textrm{cos}}
\renewcommand{\sin}{\textrm{sin}}

\renewcommand{\sinh}{\textrm{sinh}}
\renewcommand{\cosh}{\textrm{cosh}}
\renewcommand{\tanh}{\textrm{tanh}}
\newcommand{\sech}{\textrm{sech}}
\newcommand{\csch}{\textrm{csch}}

\def\exp#1{\hbox{\rm exp}\left(#1\right)}

\renewcommand{\theequation}{\arabic{section}.\arabic{equation}}
\renewcommand{\(}{\begin{equation}}
\renewcommand{\)}{end{equation} \vspace{-.05in}\linebreak}

\newcounter{saveeqn}
\newcounter{savealpheqn}

\newcommand{\alpheqn}{\setcounter{saveeqn}{\value{equation}}%
  \stepcounter{saveeqn}\setcounter{equation}{0}%
  \renewcommand{\theequation}{\mbox{\arabic{section}.\arabic{saveeqn}
\alph{equation}}}
  \renewcommand{\)}{\end{equation}}}
\def\part#1{\frac{\partial}{\partial{#1}}}%
\def\group#1{\refstepcounter{equation}\setcounter{saveeqn}
 {\value{equation}}%
  \label{#1}\setcounter{equation}{0}%
\renewcommand{\theequation}{\mbox{\arabic{section}.\arabic{saveeqn}
\alph{equation}}}
  \renewcommand{\)}{\end{equation}}}
\newcommand{\reseteqn}{\setcounter{equation}{\value{saveeqn}}%
  \renewcommand{\theequation}{\arabic{section}.\arabic{equation}}%
  \renewcommand{\)}{\end{equation}}}

\newcommand{\aalpheqn}{\setcounter{saveeqn}{\value{equation}}%
  \stepcounter{saveeqn}\setcounter{equation}{0}%
  \renewcommand{\theequation}{\mbox{
        \Alph{subsection}.\arabic{saveeqn}\alph{equation}}}
   \renewcommand{\)}{\end{equation}}}
\newcommand{\areseteqn}{\setcounter{equation}{\value{saveeqn}}%
  \renewcommand{\theequation}{\Alph{subsection}.\arabic{equation}}%
  \renewcommand{\)}{\end{equation}}}

\renewcommand{\thefootnote}{\alph{footnote}}
\renewcommand{\(}{\begin{equation}}
\renewcommand{\)}{\end{equation}}
\newcommand{\ba}{\begin{eqnarray}}
\newcommand{\ea}{\end{eqnarray}}

\renewcommand{\b}{\beta}

\newcommand{\cbp}{\mathop{\vtop{\ialign{##\crcr
   $\hfil\displaystyle{}\hfil$\crcr\noalign{\kern-13pt\nointerlineskip}
   \BIG{)}\hskip0pt\crcr\noalign{\kern3pt}}}}}
\newcommand{\pa}{\mathop{\vtop{\ialign{##\crcr

$\hfil\displaystyle{\oplus}\hfil$\crcr\noalign{\kern+1pt\nointerlineskip
}
   \hspace{.08in}$^{\alpha=0}$\hskip6pt\crcr\noalign{\kern3pt}}}}}
\renewcommand{\hsp}{,\hspace{.3in}}
\newcommand{\p}{^\prime}
\newcommand{\pp}{^{\prime\prime}}

\def\D{\ensuremath{{\cal D}}}

\catcode`\@=11
\def\vereq#1#2{\lower3pt\vbox{\baselineskip1.5pt \lineskip1.5pt
\ialign{$\m@th#1\hfill##\hfil$\crcr#2\crcr\sim\crcr}}}
\catcode`\@=12

\renewcommand{\(}{\begin{equation}}
\renewcommand{\)}{\end{equation}}

\def\pin#1{\int \frac{d#1}{2\pi}}
\def\ppin#1{\int\hspace{-17pt}\sum \frac{d#1}{2\pi}}
\def\ppink#1{\int\hspace{-17pt}\sum\frac{d^{#1}k}{(2\pi)^{#1}}}
\def\dint{\int\hspace{-12pt}\sum }
\def\pink#1{\int \frac{d^{#1}k}{(2\pi)^{#1}}}

\def\Bd#1{B^\dag_{k_{#1}}}

\def\tp#1#2#3{\hbox{\rm tan}^#1(\thet#2#3)}

\def\cc{\mathcal{C}}
\def\df{\mathcal{D}_{f}}

\def\dF{\mathcal{D}_F}

\def\I{\mathcal{I}}

\def\os{\omega_S}

\def\as{|\alpha;\sigma\rangle}
\def\asb{\langle\alpha;\sigma|}

\newcommand{\beas}{\begin{eqnarray*}}
\newcommand{\eeas}{\end{eqnarray*}}

\newcommand{\bquo}{\begin{quote}}
\newcommand{\enqu}{\end{quote}}

\newcommand{\g}{{\mathfrak g}}

\def\ch{{\mathcal{H}}}

\def\tp{{\tilde{\phi}}}
\def\ok#1{\omega_{k_{#1}}}

\def\V#1{V^{(#1)}(gf(x))}

\newcommand{\beq}{\begin{equation}}
\newcommand{\eeq}{\end{equation}}
\newcommand{\bea}{\begin{eqnarray}}
\newcommand{\eea}{\end{eqnarray}}

\newskip\humongous \humongous=0pt plus 1000pt minus 1000pt

\newif\ifdtup

\catcode`\@=11

\ifthenelse{\equal{\letter}{0}}{ 

\@addtoreset{equation}{section}
\def\theequation{\arabic{section}.\arabic{equation}}

\def\@normalsize{\@setsize\normalsize{15pt}\xiipt\@xiipt
\abovedisplayskip 14pt plus3pt minus3pt%
\belowdisplayskip \abovedisplayskip
\abovedisplayshortskip \z@ plus3pt%
\belowdisplayshortskip 7pt plus3.5pt minus0pt}

\def\small{\@setsize\small{13.6pt}\xipt\@xipt
\abovedisplayskip 13pt plus3pt minus3pt%
\belowdisplayskip \abovedisplayskip
\abovedisplayshortskip \z@ plus3pt%
\belowdisplayshortskip 7pt plus3.5pt minus0pt
\def\@listi{\parsep 4.5pt plus 2pt minus 1pt
      \itemsep \parsep
      \topsep 9pt plus 3pt minus 3pt}}

\relax

\def\section{\@startsection{section}{1}{\z@}{3.5ex plus 1ex minus  .2ex}{2.3ex plus .2ex}{\large\bf}}

\def\thesection{\arabic{section}}
\def\thesubsection{\arabic{section}.\arabic{subsection}}

\def\appendix{\setcounter{section}{0}
 \def\thesection{Appendix \Alph{section}}
 \def\thesubsection{\Alph{section}.\arabic{subsection}}
 \def\theequation{\Alph{section}.\arabic{equation}}}
\renewcommand{\theequation}{\arabic{section}.\arabic{equation}}

}{
\renewcommand{\theequation}{\arabic{equation}}

} 

\begin{document}
\def\thefootnote{\fnsymbol{footnote}}
\def\thetitle{Kink Form Factors}
\def\autone{Jarah Evslin}
\def\auttwo{Hengyuan Guo}
\def\affa{Institute of Modern Physics, NanChangLu 509, Lanzhou 730000, China}
\def\affb{University of the Chinese Academy of Sciences, YuQuanLu 19A, Beijing 100049, China}

\title{Titolo}

\ifthenelse{\equal{\pr}{1}}{
\title{\thetitle}
\author{\autone}
\affiliation {\affa}
\affiliation {\affb}

}{

\begin{center}
{\large {\bf \thetitle}}

\bigskip

\bigskip


{\large \noindent  \autone{${}^{1,2}$} }


\vskip.7cm

1) \affa\\
2) \affb\\

\end{center}

}

\begin{abstract}
\noindent
We use a recently constructed linearized soliton sector perturbation theory to calculate the form factors relevant to the elastic scattering of ultrarelativistic mesons off of nonrelativistic kinks.   Both localized kink wave packets and also delocalized momentum eigenstate kinks are considered.  In the delocalized case, the leading term is just the classical kink solution, as was found by Goldstone and Jackiw.   The leading delocalized quantum correction agrees with that found by Gervais, Jevicki and Sakita in the $\phi^4$ model and Weisz in the Sine-Gordon model.   In the case of localized kink wave packets, some corrections are found which scale with the wave packet width, and so will be relevant for the coherent scattering of mesons off of kink wave packets.

\end{abstract}

%
\setcounter{footnote}{0}
\renewcommand{\thefootnote}{\arabic{footnote}}

\ifthenelse{\equal{\pr}{1}}
{
\maketitle
}{}

\section{Introduction}
Consider a classical field theory whose interactions are described by a coupling $g$ with dimensions of [action]${}^{-1/2}$.  Let it have one, or more, homogeneous field configurations which minimize the energy.  In a corresponding quantum theory, if $g\sqrt{\hbar}$ is small, generally there will be a quantum state, called a vacuum, corresponding to each minimum of the classical energy.  There will also be a Fock space of perturbative excitations above this vacuum.  We will refer to all of these excitations, with an excitation number of order unity (and not $1/(g\sqrt{\hbar})$, for example) as the vacuum sector.

If there is a localized stationary classical solution, such as a topological soliton, then in many cases of interest there will be a quantum state corresponding to the classical solution\footnote{This state is not guaranteed to be a Hamiltonian eigenstate.  There may be quantum mechanical instabilities \cite{weigelstab} and {\it{vice versa}} a classically unstable solution may be stable in the quantum theory \cite{delfino,davies}.}.  Such a  state does not lie in the vacuum sector.  It may also be excited, or more precisely its normal modes may be excited.  We refer to states in which of order unity normal modes are excited as the soliton sector.

In the case of scalar theories in (1+1)-dimensions, if the potential contains degenerate minima then there will be classical kink solutions.  Under certain, rather special, conditions these correspond to Hamiltonian eigenstates in the quantum theory.  The spectrum of the quantum kink sector was found at one loop in Ref.~\cite{dhn2}.  There are now many powerful methods available at one loop \cite{rajaraman75,goldhaber04,weigrev}.  Progress beyond one loop is complicated by the continuously degenerate soliton spectrum corresponding to the choice of position of the soliton.  

This problem is usually treated using the collective coordinate approach of Ref.~\cite{gjscc}.  In the collective coordinate approach, the kink position is promoted to an operator.  One then isolates its conjugate momentum and performs a nonlinear canonical transformation to disentangle these two operators from the other operators in the theory.  This nonlinear transformation is already rather complicated in the classical theory, but in the quantum theory it also leads to additional interaction terms in the Hamiltonian \cite{gj76}.  In principle this method allows any problem to be solved.  However it is prohibitively difficult.  As a result the most basic quantity that may be computed beyond one loop, the two-loop kink rest mass, has only been computed using collective coordinates in cases where it was already known as a result of integrability \cite{vega,verwaest} or supersymmetry \cite{shif99}.  At one loop it has had more applications.  For example it has been applied to compute form factors of the $\phi^4$ theory \cite{gjscc}, although counterterms needed to render the answer finite were not included.

Recently a less powerful method has been proposed.  A base point is chosen in the space of kink locations, and all fields are expanded with respect to the normal modes at this base point.  In particular, the resulting zero mode agrees with the collective coordinate at linear order, but differs at higher orders.  The degeneracy problem is then resolved by fixing the momentum of the desired state in perturbation theory.  This is a series of linear constraints, and so the nonlinear canonical transformation is avoided.  The price to be paid is that this series only converges for kink positions sufficiently close to the base point, and so one cannot consider a coherent superposition of well-separated kinks.  Nonetheless, for problems such as finding the energy spectrum, this light-weight method is sufficient.

So far this linearized soliton sector perturbation theory has been used to calculate the two-loop masses of kink ground states \cite{me2loop,mephi4}, the leading corrections to the energy required to excite kink shape modes and continuum excitations \cite{me2loopex} and the instantaneous acceleration of a kink in the presence of an impurity \cite{chris}.   For this last calculation, it was necessary to consider the full position-dependence of the kink wave function.  Due to the convergence issues described above, this restricted the range of validity to localized wave packets \cite{meimpulso}, which anyway are the ones of interest to scattering off of fixed impurities.  However in many cases of interest, such as the high energy scattering of skyrmion \cite{skyrme,wittenskyrme,smorg} models of baryons, to a good approximation the soliton wave function is a plane wave and not a localized wave packet.

Our next goal is to apply this linearized perturbation theory to the scattering of nonrelativistic kinks with ultrarelativistic mesons \cite{faddeev77,lowe79,parm87,swanson88,uehara91,abdel11}.  In this note, we will calculate the form factors relevant to elastic kink-meson scattering\footnote{As a result of our perturbative expansion we can only consider nonrelativistic kinks.  However recently an approach to form factors involving relativistic kinks has been presented in Ref.~\cite{andyff2}.}.  These are Schrodinger picture form factors, which are the matrix element of a scalar field sandwiched between two wave packets of ground state kinks, all at equal time.   This matrix element is therefore an amplitude for the instantaneous emission or absorption of a meson by a ground state kink.  Such a process of course cannot be on shell, but a pair of such matrix elements appears, for example, in elastic scattering.   The methods used here can be straightforwardly extended to excited kink states, which will allow an application to inelastic scattering, radiative and nonradiative meson absorption and spontaneous or induced meson emission.    We intend to treat these specific processes in future works.   A generalization to matrix elements involving multiple meson fields is also straightforward, and will be relevant to the above processes at higher orders and to other processes, such as decays of multiple shape mode excitations, as well as processes in more complicated models \cite{loginov22}.

We begin in Sec.~\ref{revsez} with a review of linearized soliton sector perturbation theory, applied to kinks in (1+1)-dimensional scalar field theories.  Then, in the case of localized kinks, in Sec.~\ref{classsez} we define the form factors that we will compute and determine the leading contribution.  Although we consider localized wave packets, our leading contribution is just the Fourier transform of the classical solution, in agreement with the case of plane waves of kinks in Ref.~\cite{gj75}.  Next in Sec.~\ref{quantsez} we compute the leading corrections.  These correspond to higher order corrections to the boost operator, to the ground state and to the normalization.  In Sec.~\ref{delsez}, we argue that the form factor of a delocalized kink is given by a subset of the terms that we found for the localized kinks, at least when the momentum transfer is much greater than the meson mass.  Finally, in Sec.~\ref{sgsez}, we find the leading correction to the form factor of the Sine-Gordon soliton, and show that it agrees with the answer that was obtained in Refs.~\cite{weisz77,bab01} using the integrability of that model.

\section{Linearized Soliton Sector Perturbation Theory} \label{revsez}

\subsection{The Kink Hamiltonian and Hilbert Space}

We will be interested in a (1+1)-dimensional theory of a scalar field $\phi(x)$ with canonical momentum $\pi(x)$ and a degenerate potential $V$ defined by the Hamiltonian
\bea
H[\pi(x),\phi(x)]&=&\int dx :\ch(\pi(x),\phi(x)):_a\\
\ch(\pi(x),\phi(x))&=&\frac{1}{2} \left(\pi^2(x)+\left(\partial_x\phi(x)\right)^2\right)+\frac{1}{g^2}V(g\phi(x))\nonumber
\eea
where the normal-ordering prescription $::_a$ will be defined below.   We will expand the Hamiltonian, our states and our energies in the small coupling $g$, where $\hbar=1$.  It is understood that all fields and states in this note are in the Schrodinger picture.

Consider a stationary kink solution of the classical equations of motion
\beq
\phi(x,t)=f(x)\hsp
f\pp(x)=\frac{1}{g}\V{1} \label{cleq}
\eeq
where we have defined
\beq
\V{n}=\frac{\partial^n}{\partial(g\phi(x))^n}V(g\phi(x))|_{\phi(x)=f(x)}.
\eeq
The classical equations of motion are nonlinear, although they linearize for small perturbations.  $f(x)$ is a soliton solution however, and so by definition is sufficiently large that it is well into the nonlinear regime.   This suggests that the quantum kink cannot be studied in perturbation theory.

Our goal is to study small perturbations about the kink in perturbation theory.  Classically, small excitations of the kink correspond to a classical field $\phi(x,t)$ which is equal to $f(x)$ plus a small perturbation.   Thus these small perturbations, $\phi(x,t)-f(x)$, satisfy a linear equation (\ref{sl}) and can be studied in perturbation theory.  The first step in this approach is to write a Hamiltonian for these small perturbations.

To do this, we define the unitary displacement operator
\beq
\df={\rm{exp}}\left(-i\int dx f(x)\pi(x)\right) \label{df}
\eeq
which, for any normal ordering prescription and any functional $F[\phi(x),\pi(x)]$ transforms
\beq
:F[\phi(x),\pi(x)]:\df=\df:F[\phi(x)+f(x),\pi(x)]:. \label{dfa}
\eeq
In other words, when on the left hand side the argument $\phi(x)$ is a small perturbation of $f(x)$, on the right hand side the corresponding $\phi(x)$ is small and so may be treated perturbatively.  We use this operator to transform the defining regularized Hamiltonian $H$, momentum $P$ and boost operators $\Lambda$, which are nonperturbative when applied to the kink sector, into the kink Hamiltonian, momentum and boost operators
\beq
H\p=\df^\dag H\df\hsp
P\p=\df^\dag P\df\hsp
\Lambda\p=\df^\dag \Lambda\df. \label{hp}
\eeq
We let these operators act not on the original, defining Hilbert space, but rather on the kink Hilbert space, which is related to the defining Hilbert space by the action of $\df$.  

How is this useful?  Imagine that we find an eigenstate $\vac$ of the kink Hamiltonian $H\p$, with eigenvector $Q$.  For kink sector states, we have argued that such eigenstates can be found in perturbation theory.  Then, $\df\vac$ will be an eigenstate of the defining Hamiltonian $H$ with the same eigenvector $Q$
\beq
H\p\vac=Q\vac \Rightarrow H\df\vac=Q\df\vac.
\eeq
Thus we have solved the $H$ eigenvalue problem, which we would expect to be nonperturbative, by working in the kink Hilbert space, where it is perturbative.  

This is just the quantum version of the classical physics procedure of first performing a transformation $\phi(x)\rightarrow\phi(x)-f(x)$ of the fields so that the fields are small, and then linearizing about these small field values.  Historically, beginning with Ref.~\cite{dhn2}, the kink Hamiltonian was constructed via precisely this transformation.  However it was discovered in Ref.~\cite{rebhan} that this transformation does not always commute with the regularization which is required to construct the quantum theory.  As a result the eigenvalue of $H\p$ did not produce the correct mass, which is defined to be the eigenvalue of $H$.  This problem is resolved in our formulation, as $H\p$ and the regularized $H$ are related by a similarity transformation (\ref{hp}) and so their eigenvalues necessarily agree.  More concretely, whereas traditionally authors first constructed the kink Hamiltonian, then regularized both the defining and also the kink Hamiltonians, and then introduced an {\it ad hoc} matching condition on these regulators, we first regularize the defining Hamiltonian and then use it to construct the kink Hamiltonian, which is therefore created already regularized.

\subsection{Decomposing the Fields into Plane Waves and Normal Modes}

Small, constant frequency perturbations about the vacuum in classical field theory are plane waves.  As a result, the first step in a perturbative treatment of the vacuum sector of the quantum theory is the decomposition of the fields into plane waves
\beq
\tp_p=\int dx \phi(x)e^{ipx}\hsp
\tilde{\pi}_p=\int dx \pi(x)e^{ipx} \label{pdec}
\eeq
which in turn can be decomposed into creation and annihilation operators
\beq
A^\dag_p=\frac{\tp_p}{2}-i\frac{\tilde{\pi}_p}{2\omega_p}\hsp
\frac{A_{-p}}{2\omega_p}=\frac{\tp_p}{2}+i\frac{\tilde{\pi}_p}{2\omega_p}\hsp
\omega_p=\sqrt{m^2+p^2}.
\eeq
Here $2\omega_p A^\dag_p$ is the Hermitian conjugate of $A_p$ and $m$ is the second derivative of $V$ at the classical minimum of the potential.  If the two classical minima on opposite sides of the kink have different second derivatives, then the kink will accelerate \cite{tstabile,wstabile} and there is no corresponding Hamiltonian eigenstate, so we are not interested in such cases.

On the other hand, small constant frequency perturbations about the kink in classical field theory are normal modes $\g$ which solve
\beq
\V2\g(x)=\omega^2\g(x)+\g\pp(x). \label{sl}
\eeq
More precisely there is a zero mode $\g_B(x)$ with frequency $\omega_B=0$, a continuum of modes $\g_k(x)$ for all real $k$ with frequencies $\ok{}=\sqrt{m^2+k^2}$ and sometimes there are discrete shape modes $\g_S(x)$ with $0<\omega_S<m$.  The discrete modes are taken to be real, while for the continuum modes we impose $\g_k^*(x)=\g_{-k}(x)$.  In the Schrodinger picture, fields are independent of time and so we may decompose them in any basis of functions.  The normal modes solve a Sturm-Liouville equation (\ref{sl}) and so provide a basis. Therefore, in the kink sector it is convenient to decompose the fields in the normal mode basis
\beq
\tp_k=\int dx \phi(x) \g^*_k(x)\hsp
\tilde{\pi}_k=\int dx \pi(x) \g^*_k(x)
\eeq
where $k$ is a real number for continuum modes but also runs over all discrete indices $S$ and $B$.  We write $\phi_0$ and $\pi_0$, and not $\tp_B$ and $\tilde{\pi}_B$, for the zero modes.  To avoid confusion between this decomposition and the plane wave decomposition (\ref{pdec}), we use the letters $p$ and $q$ exclusively for plane wave momenta and $k$ exclusively for normal modes.  

All of the modes except for the zero modes may be rewritten in terms of annihilation and creation operators
\beq
B^\dag_k=\frac{\tp_k}{2}-i\frac{\tilde{\pi}_k}{2\omega_k}\hsp
\frac{B_{-k}}{2\omega_k}=\frac{\tp_k}{2}+i\frac{\tilde{\pi}_k}{2\omega_k}
\eeq
where $2\omega_p B^\dag_k$ is the Hermitian conjugate of $B_k$.  

We normalize the normal modes so that
\beq
\int dx |{\g}_{B}(x)|^2=1,\
\int dx {\g}_{k_1} (x) {\g}^*_{k_2}(x)=2\pi \delta(k_1-k_2),\ 
\int dx {\g}_{S_1}(x){\g}_{S_2}(x)=\delta_{S_1S_2}.
\eeq\
The corresponding completeness relation is 
\beq
{\g}_B(x){\g}_B(y)+\ppin{k}{\g}_k(x){\g}^*_{k}(y)=\delta(x-y). \label{comp}
\eeq
Here we have introduced $\dint$ which integrates over continuum modes and sums over shape modes, but does not include zero modes.  We choose the sign of $\g_B(x)$ so that
\beq
f\p(x)=\sqrt{Q_0} \g_B(x). \label{fg}
\eeq

\begin{table}
\begin{tabular}{|l|l|l|} 
\hline
Name&Basis&Algebra\\
\hline\hline
Position space fields&$\phi(x),\ \pi(x)$&$[\phi(x),\pi(y)]=i\delta(x-y)$\\
\hline
Momentum space fields&$\tp_p,\ \tilde{\pi}_p $&$[\tp_p,\tilde{\pi}_q]=2\pi i \delta(p+q)$\\
\hline
Normal mode fields&$\tp_k,\tilde{\pi}_k,$&$[\tp_{k_1},\tilde{\pi}_{k_2}]=2\pi i \delta(k_1+k_2),$\\
&$\tp_{S},\tilde{\pi}_S,\phi_0,\pi_0  $&$[\tp_{S_1},\tilde{\pi}_{S_2}]=i\delta_{S_1S_2},\ [\phi_0,\pi_0]=i$\\
\hline
Plane wave operators&$A^\dag_p,\ A_p$&$[A_p,A^\dag_q]=2\pi\delta(p-q)$\\
\hline
Normal mode operators&$B^\dag_k,B_k,$&$[B_{k_1},B^\dag_{k_2}]=2\pi\delta(k_1-k_2) $\\
&$B^\dag_S,B_S,\phi_0,\pi_0$&$[B_{S_1},B^\dag_{S_2}]=\delta_{S_1S_2},\ [\phi_0,\pi_0]=i$\\
\hline
\end{tabular} 
\caption{Bases of operator algebra}\label{bastab}
\end{table}

We have thus found five bases of our algebra of operators, listed in Table~\ref{bastab}.   Any operator may be expanded in any basis and there are Bogoliubov transformations which let one change one basis to another.  In the plane wave and normal mode creation and annihilation operator bases, there are corresponding normal ordering prescriptions.  These are defined as follows.  The operator $:O:_a$ or $:O:_b$ is called plane wave or normal mode normal ordered respectively if, when expressed in the plane wave or normal mode basis, all $A^\dag$ or all $B^\dag$ and $\phi_0$ appear on the left. 

\subsection{The Kink Hamiltonian}

What is the kink Hamiltonian?  From Eq.~(\ref{dfa}) one can see that it is equal to
\beq
H\p[\pi(x),\phi(x)]=\int dx :\ch\p(\pi(x),\phi(x)):_a\hsp
\ch\p(\pi(x),\phi(x))=\ch(\pi(x),\phi(x)+f(x)). \label{hkd}
\eeq
Now let us decompose it
\beq
H_n=\int dx \ch_n
\eeq
where $\ch_n$ contains all terms in $\ch\p$ which, when plane wave normal ordered, contain $n$ factors of $\phi(x)$ and $\pi(x)$.  The terms are easily evaluated.  The zeroeth is just the mass $Q_0$ of the classical kink
\beq
H_0=Q_0.
\eeq
The first, $H_1$, vanishes.  The free part of the theory is
\beq
\ch_2(x)=\frac{1}{2}\left[
:\pi^2(x):_a+:\left(\partial_x\phi(x)\right)^2:_a+\V2 :\phi^2(x):_a
\right]. \label{fh}
\eeq
The interaction terms are
\beq
\ch_{n>2}(x)=\frac{g^{n-2}}{n!}\V{n} :\phi^n(x):_a. \label{hint}
\eeq

The free part of the Hamiltonian in Eq.~(\ref{fh}) is plane wave normal ordered.  It looks like a usual free Hamiltonian except for the position-dependent mass term.  To find its eigenstates, it is convenient to normal mode normal order it.  This yields \cite{mekink}
\beq 
H_2=Q_1+\frac{\pi_0^2}{2}+\os B^\dag_SB_S+\ppin{k}\ok{} B^\dag_kB_k \label{h2p}
\eeq
where for concreteness we have considered a single shape mode.  Here $Q_1$ is equal to the one-loop correction to the kink mass, given in the Cahill-Comtets-Glauber \cite{cahill76} form.   This Hamiltonian is a sum of free quantum mechanical Hamiltonians.  The first is the kinetic energy of a free particle, representing the kink center of mass.  More precisely, $\sqrt{Q_0}\pi_0$ is the momentum operator for the kink center of mass, and so $\phi_0/\sqrt{Q_0}$ is the position operator.  The other terms are quantum harmonic oscillators for the normal modes.

\subsection{The Kink Ground State in Perturbation Theory}

In the kink Hilbert space, we denote the kink ground state by $\vac$.  Recall that it is an eigenvalue of the kink Hamiltonian $H\p$ with eigenvalue $Q$.  To find $\vac$ in perturbation theory, we expand
\beq
\vac=\sum_{i=0}^\infty \vac_i\hsp Q=\sum_{i=0}^\infty Q_i
\eeq
where $\vac_i$ is suppressed with respect to $\vac_0$ by $g^i$ and $Q_i$ is of order $O(mg^{2i-2})$.  

The leading terms in this expansion solve the eigenvalue problem for $H_0+H_2$
\beq
(H_0+H_2)\vac_0=(Q_0+Q_1)\vac_0.
\eeq
Recalling that $H_0=Q_0$, these terms can be removed from the equation.  Then Eq.~(\ref{h2p}) implies that $\vac_0$ is the ground state of all of the oscillators, with the center of mass momentum turned off
\beq
\pi_0\vac_0=B_S\vac_0=B_k\vac_0=0.
\eeq
The states $\vac_1$ and $\vac_2$ were found in Ref.~\cite{me2loop} while the corresponding states for excited kinks were given in Ref.~\cite{me2loopex}.  The center of mass motion was considered in Ref.~\cite{meimpulso}, where the operator $\Lambda\p$ was constructed which boosts kinks.  It was expanded in operators $\Lambda\p_i$ each of which contain $i$ factors of the fields when plane wave normal ordered.

\section{The Leading Order Form Factor} \label{classsez}

\subsection{Definitions}

In the kink sector Hilbert space, consider the wave packet \cite{meimpulso}
\beq
\as=\frac{\sqrt{\mathcal{N}}}{(2\pi)^{1/4}\sqrt{\sigma}}e^{-\frac{\phi_0^2}{4\sigma^2}}e^{i\alpha\Lambda\p}\vac \label{sig}
\eeq
where the normalization constant $\mathcal{N}$ is chosen so that
\beq
\langle\alpha;\sigma\as=1. \label{norm}
\eeq
Recalling that $\phi_0/\sqrt{Q_0}$ is the position operator of the center of mass of the kink, $\sigma/\sqrt{Q_0}$ is the position-space width of the corresponding wave packet.  The state $\as$ in the kink Hilbert space corresponds to the state $\df\as$ in the defining Hilbert space, which is a kink wave packet with expected rapidity $\alpha$.  We define the form factor $\tilde{\ff}_q$ and its Fourier transform $\ff(z)$ by
\beq
\tilde{\ff}_q=\langle 0;\sigma|\df^\dag \tp_q \df\as=\int dz \ff(z) e^{iqz}. \label{ffdef}
\eeq
When $q=-Q_0 \alpha$, the expected momentum of $\as$ cancels that of $\tp_q$.

There are four important dimensionless parameters.  The first is the coupling $g$.  The second is the expected rapidity $\alpha$ of the kink.  Both of these need to be taken small in our semiclassical expansion.  The third is $q/m$, which is large for an ultrarelativistic meson.  The momentum space width of our wave packet is larger than $m$ \cite{meimpulso} and so only for ultrarelativistic mesons is the expected momentum much greater than the momentum spread.  The last parameter is $\sigma\sqrt{m}$, which is the wave packet width in units of the meson mass, or intuitively in units of the width of the classical kink solution.  This last parameter is, at this point, unconstrained.  However, the methodology of Ref.~\cite{me2loop} computes the states as an expansion in, among other things, $gm\phi_0^2$ and so this perturbative parameter is only small if
\beq
\sigma\sqrt{m}<<\frac{1}{\sqrt{g}}. \label{smax}
\eeq 

\subsection{Leading Order Form Factor}

At leading order in the coupling $g$, the wave packets are
\beq
\as_0=\frac{1}{(2\pi)^{1/4}\sqrt{\sigma}}e^{-\frac{\phi_0^2}{4\sigma^2}}e^{i\alpha\Lambda\p_1}\vac_0 \label{sig0}
\eeq
and the form factor is
\beq
\tilde{\ff}_{{\rm{tree}},q}={}_0\langle 0;\sigma|\df^\dag \tp_q \df \as_0.
\eeq
In Ref.~\cite{meimpulso} we found that the leading order boost operator is
\beq
\Lambda\p_1=-\sqrt{Q_0}\phi_0
\eeq
and so
\bea
\tilde{\ff}_{{\rm{tree}},q}&=&\int dx e^{iqx}{}_0\langle 0;\sigma|\df^\dag \phi(x) \df\as_0=\int dx e^{iqx}{}_0\langle 0;\sigma|(\phi(x)+f(x))\as_0\label{ff00}\\
&=&\int dx e^{iqx}{}_0\langle 0;\sigma|(\phi_0 \g_B(x)+f(x))\as_0\nonumber\\
&=&\frac{1}{\sigma\sqrt{2\pi}}\int dx e^{iqx}{}_0\langle 0|e^{-\frac{\phi_0^2}{4\sigma^2}}\left(\phi_0 \frac{f\p(x)}{\sqrt{Q_0}}+f(x)\right)e^{-\frac{\phi_0^2}{4\sigma^2}-i\alpha\sqrt{Q_0}\phi_0}\vac_0\nonumber\\
&=&\frac{1}{\sigma\sqrt{2\pi}}\int dx e^{iqx}\int dy e^{-\frac{y^2}{2\sigma^2}-i\alpha\sqrt{Q_0}y}\left(f(x)+\frac{y}{\sqrt{Q_0}}f\p(x)
\right). \nonumber
\eea

\begin{table}
\begin{tabular}{|l|l|l|} 
\hline
Name&Definition&Interpretation\\
\hline\hline
$x$&&Position coordinate  in the laboratory frame\\
\hline
$-y/\sqrt{Q_0}$&$\phi_0|y\rangle_0=y|y\rangle_0$&Position of the kink center of mass in the lab frame\\
\hline
$z$&$z=x+\frac{y}{\sqrt{Q_0}}$&Position coordinate in the kink center of mass frame\\
\hline
\end{tabular} 
\caption{Coordinates}\label{cotab}
\end{table}

In the last step we decomposed $\vac_0$ into the states $|y\rangle_0$ defined by
\beq
\phi_0|y\rangle_0=y|y\rangle_0\hsp
B_k|y\rangle_0=0.
\eeq
The decomposition is
\beq
\vac_0=\int dy |y\rangle
\eeq
and we recall \cite{meimpulso} that $-y/\sqrt{Q_0}$ is the position of the kink.  Therefore 
\beq
z=x+\frac{y}{\sqrt{Q_0}}
\eeq
is the position coordinate relative to the kink.  This situation is summarized in Table~\ref{cotab}.

Rewriting (\ref{ff00}) in terms of $z$ one finds
\bea
\tilde{\ff}_{{\rm{tree}},q}\,\,&=&\,\,
\int dz e^{iqz}\int dy \frac{e^{-\frac{y^2}{2\sigma^2}-i(Q_0\alpha+q)y/\sqrt{Q_0}}}{\sigma\sqrt{2\pi}}\left(f\left(z-\frac{y}{\sqrt{Q_0}}\right)+\frac{y}{\sqrt{Q_0}}f\p\left(z-\frac{y}{\sqrt{Q_0}}\right)\right)\label{esp}\\
&=&\,\,
\int dz e^{iqz}\int dy \frac{e^{-\frac{y^2}{2\sigma^2}-i(Q_0\alpha+q)y/\sqrt{Q_0}}}{\sigma\sqrt{2\pi}}\left(f(z)-\sum_{j=2}^{\infty} \frac{1}{j!}\left(\frac{y}{\sqrt{Q_0}}\right)^j  f^{(j)}\left(z-\frac{y}{\sqrt{Q_0}}\right)\right).\nonumber
\eea
Recalling that $1/\sqrt{Q_0}$ is of order $g$, one sees that the sum over $j$ is a perturbative expansion in the coupling.  The leading term is
\beq
\tilde{\ff}_{0,q}=
\frac{1}{\sigma\sqrt{2\pi}}\int dz e^{iqz}f(z)\int dy e^{-\frac{y^2}{2\sigma^2}-i(Q_0\alpha+q)y/\sqrt{Q_0}}=\int dz e^{iqz}f(z)e^{-\frac{\sigma^2\left(Q_0\alpha+q\right)^2}{2Q_0}}.
\eeq
We would like to interpret the expression on the right as the Fourier transformed form factor, but if $q\neq-Q_0\alpha$ it depends on q.    What does this mean?

If $q=-Q_0\alpha$, then the expectation value of the momentum of our wave packet $\as_0$ is equal and opposite to that of the operator $\tp_q$.  These theories are translation-invariant and so momentum is conserved.  This means that if we decompose the definition of the form factor in terms of momentum eigenstates, then only kets with momentum precisely $q$ less than bras will contribute.  At $q=-Q_0\alpha$, the peaks of $\as$ and $\langle \sigma;0|$ have momentum $Q_0\alpha$ and $0$ respectively and so satisfy this condition.  In the limit $\sigma\rightarrow\infty$, which is beyond the validity of our perturbative expansion of the states but nonetheless well-defined, the wave packet becomes a plane wave with momentum $Q_0\alpha$ and so the form factor is only nonvanishing at $q=-Q_0\alpha$, whereas for finite $\sigma$ the momentum spread is of order $\sqrt{Q_0}/\sigma$.   Thus in any case, a nontrivial contribution only occurs for $q$ close to $-Q_0\alpha$, and one expects the largest form factor at $q=-Q_0\alpha$.

In the case $q=-Q_0\alpha$, the form factor simplifies to
\beq
\tilde{\ff}_{0,q=-Q_0\alpha}=\int dz e^{iqz}f(z)
\eeq
and so its Fourier transform is just the classical solution
\beq
\ff_0(z)=\pin{q} e^{-iqz}\tilde{\ff}_{0,q=-Q_0\alpha}=f(z)
\eeq
as was shown in Ref.~\cite{gj75} in the case $\sigma=\infty$.

However, the finite spread in the momentum of our wave packet means that we may also consider the form factor off of the momentum peak of the wave function.  Let us consider a fixed momentum offset
\beq
\epsilon=Q_0\alpha+q. \label{ep}
\eeq
Obviously
\beq
\tilde{\ff}_{0,q=\epsilon-Q_0\alpha}=e^{-\frac{\sigma^2\epsilon^2}{2Q_0}}\int dz e^{iqz}f(z).
\eeq
Now, one may calculate the Fourier transform of the form factor with $\epsilon$ held fixed
\beq
\ff_{0,\epsilon}(z)=\pin{q} e^{-iqz}\tilde{\ff}_{0,q=\epsilon-Q_0\alpha}=e^{-\frac{\sigma^2\epsilon^2}{2Q_0}}f(z). \label{ff0}
\eeq
This is easy to interpret.  It means that the amplitude for any process where the wave packet creates or destroys a scalar off of its momentum peak is suppressed by a Gaussian equal to the Fourier transform of the wave packet, in other words, by the momentum-space wave function of the wave packet.




\section{Corrections} \label{quantsez}

In this section we will systematically study the dominant corrections to the form factor (\ref{ff0}).  These include all of the corrections up to linear order in the coupling $g$.  This calculation was begun in Ref.~\cite{gervaisff75}, in the case of a delocalized kink, and we will see that their result appears as one of the corrections below.

\subsection{The Second Derivative of the Classical Solution}

Recall that $\tilde{\ff}_{{\rm{tree}},q}$ in Eq.~(\ref{esp}) contained a power series in the classical kink solution $f$.  The dominant term $\tilde{\ff}_{0,q}$ was the constant term in this power series.  There was no linear term, but the second derivative term was nonvanishing.  By shifting the terms at three derivatives and higher, the second derivative term in (\ref{esp}) may be written
\bea
\tilde{\cc}_{1,q}&=&-\frac{1}{2}
\int dz e^{iqz}\int dy \frac{e^{-\frac{y^2}{2\sigma^2}-i(Q_0\alpha+q)y/\sqrt{Q_0}}}{\sigma\sqrt{2\pi}}\left(\frac{y}{\sqrt{Q_0}}\right)^2  f\pp(z)\\
&=&\int dz e^{iqz}\left[
-\frac{f\pp(z)}{2Q_0}\sigma^2\left(1-\frac{\sigma^2\left(Q_0\alpha+q\right)^2}{Q_0}\right)e^{-\frac{\sigma^2\left(Q_0\alpha+q\right)^2}{2Q_0}}
\right].\nonumber
\eea
Again we can define $\epsilon$ as in (\ref{ep}) as the momentum distance from the peak of the wave packet.  Then
\beq
\tilde{\cc}_{1,q}=\int dz e^{iqz}\cc_{1,\epsilon}(z)
\hsp
\cc_{1,\epsilon}(z)=-\frac{f\pp(z)}{2Q_0}\sigma^2\left(1-\frac{\sigma^2\epsilon^2}{Q_0}\right)e^{-\frac{\sigma^2\epsilon^2}{2Q_0}}. \label{c10}
\eeq
As in the case of the leading order term, we took the Fourier transform with $\epsilon$ fixed.  At $\epsilon=0$, this simplifies to
\beq
\cc_{1}(z)=-\frac{f\pp(z)}{2Q_0}\sigma^2.
\eeq
We see the suppression off of the momentum space wave packet peak is no longer just the momentum space wave function, there is an additional factor.  However the standard deviation of the momentum distribution is still of order the width of the momentum space wave packet.


How subdominant is this correction?  Let us fix $\epsilon=0$ for concreteness.  Recall that $f$ is of order $1/g$ and so the leading order form factor is of order $1/g$
\beq
\ff_{0}(z)=f(z)\sim O(1/g).
\eeq
On the other hand, $f\pp$ is of order $m^2/g$ and $Q_0$ is of order $m/g^2$.  Therefore the correction $\cc_{1}(z)$ is of order 
\beq
\cc_{1}(z)\sim O(g(\sigma^2m)).
\eeq
Therefore it is suppressed by order $g^2(\sigma^2m)$.  According to (\ref{smax}), our perturbative expansion is only valid when $g\sigma^2m<<1$ and so $g^2(\sigma^2m)$ is also very small.  Thus this contribution is indeed subleading.

\subsection{Leading Correction to the Boost Operator}

Recall that the form factor is defined in Eq.~(\ref{ffdef}) in terms of the state $\as$.  However in the previous subsection we only considered the leading order state $\as_0$.  We now want to include the leading corrections to this state.  There are two kinks of corrections.  In the next subsection, we will consider corrections to the zero-momentum kink ground state $\vac_0$.  In this subsection, we will instead consider corrections to the boost operator $\Lambda\p$.  Including the leading correction to the boost operator, our state is
\beq
\as_0+\as_{1,0}=\frac{1}{(2\pi)^{1/4}\sqrt{\sigma}}e^{-\frac{\phi_0^2}{4\sigma^2}}e^{i\alpha(\Lambda\p_1+\Lambda\p_2)}\vac_0 
\eeq
where the leading correction to the boost operator is \cite{meimpulso}
\beq
\Lambda\p_2=\ppink{2}\frac{\Delta_{k_1k_2}}{\omega_{k_2}^2-\omega_{k_1}^2}:\left(\pi_{k_1}\pi_{k_2}+\omega_{k_1}^2\phi_{k_1}\phi_{k_2}\right):_b+\ppin{k}\Delta_{B k}\left(\frac{2}{\omega^2_k}\pi_{0}\pi_{k}+ \phi_{0}\phi_{k}\right)  \label{l2}
\eeq
and we have defined the symbol
\beq
\Delta_{ij}=\int dx \g_i(x) \g\p_j(x).
\eeq

If we separate the boosts, to create a factorized product $e^{i\alpha\Lambda\p_1}e^{i\alpha\Lambda\p_2}$, then we must also include terms with commutators of $\Lambda\p_1$ and $\Lambda\p_2$.  These terms are all of quadratic order or higher in $\alpha$, and so we will drop them.  Once the boost operator is factorized, we will further consider only the linear order in the second exponential, as higher orders will again be suppressed by powers of $\alpha$.  Thus we have approximated
\beq
e^{i\alpha(\Lambda\p_1+\Lambda\p_2)}\sim e^{i\alpha\Lambda\p_1}\left(1+i\alpha\Lambda\p_2\right).
\eeq
Of the three terms in Eq.~(\ref{l2}), we may now ignore the third because $\pi_0$ annihilates $\vac_0$.  Furthermore, as a result of the normal mode normal ordering, the first two terms create two normal modes and so their matrix elements with $\phi(x)$ and $f(x)$ vanish, as these destroy at most one or zero modes respectively.  Therefore only the last term will contribute.   Thus our approximation becomes
\beq
\as_0+\as_{1,0}=\frac{1}{(2\pi)^{1/4}\sqrt{\sigma}}e^{-\frac{\phi_0^2}{4\sigma^2}-i\alpha\sqrt{Q_0}\phi_0}\left(1+i\alpha\phi_{0}\ppin{k}\Delta_{B k}\phi_{k}\right)\vac_0
\eeq
and so the correction to the state is
\beq
\as_{1,0}=\frac{i\alpha\phi_{0}}{(2\pi)^{1/4}\sqrt{\sigma}}e^{-\frac{\phi_0^2}{4\sigma^2}-i\alpha\sqrt{Q_0}\phi_0}\ppin{k}\Delta_{Bk}B^\dag_{k}\vac_0.
\eeq
On the other hand, there is no correction to the bra in (\ref{ffdef}) because it is not boosted, and so $\alpha=0$.

Altogether, our correction to the form factor is
\bea
\tilde{\cc}_{2,q}&=&{}_0\langle 0;\sigma|\df^\dag \tp_q \df \as_{1,0}={}_0\langle 0;\sigma| \tp_q  \as_{1,0}\\
&=&\int dx e^{iqx}{}_0\langle 0;\sigma| \phi(x) \as_{1,0}
=\int dx e^{iqx}\ppin{k} \frac{\g_{k}(x)}{2\ok{}} {}_0\langle 0;\sigma| B_{-k} \as_{1,0}
\nonumber\\
&=&\int dx e^{iqx}\frac{i\alpha}{\sigma\sqrt{2\pi}}\left[\ppin{k} \frac{\g_{-k}(x)\Delta_{Bk}}{2\ok{}}\right]\int dy y e^{-\frac{y^2}{2\sigma^2}-i\alpha\sqrt{Q_0}y}.
\nonumber
\eea
Again we replace the coordinate $x$ with respect to the laboratory with the coordinate $z$ with respect to the kink
\bea
\tilde{\cc}_{2,q}&=&\int dz e^{iqz}\frac{i\alpha}{\sigma\sqrt{2\pi}}\int dy \left[\ppin{k} \frac{\g_{-k}\left(z-\frac{y}{\sqrt{Q_0}}\right)\Delta_{Bk}}{2\ok{}}\right] y e^{-\frac{y^2}{2\sigma^2}-i(Q_0\alpha+q)y/\sqrt{Q_0}}.
\eea

To better understand this term we will simplify it by expanding the $\g_{-k}$ term in $y$, and keeping only the constant and linear terms.  For simplicity, we will continue to refer to this approximation as $\cc_{2,q}$.  This yields
\bea
\tilde{\cc}_{2,q}&=&\int dz e^{iqz}
\frac{i\alpha}{\sigma\sqrt{2\pi}}\ppin{k} \frac{\Delta_{Bk}}{2\ok{}}\left[\g_{-k}(z)  \int dy  y e^{-\frac{y^2}{2\sigma^2}-i(Q_0\alpha+q)y/\sqrt{Q_0}}
\right.\\
&&\left.- \frac{\g\p_{-k}(z)}{\sqrt{Q_0}}\int dy  y^2 e^{-\frac{y^2}{2\sigma^2}-i(Q_0\alpha+q)y/\sqrt{Q_0}}\right]\nonumber\\
&=&\int dz e^{iqz}\left[
-i\alpha \sigma^2 e^{-\frac{\sigma^2\left(Q_0\alpha+q\right)^2}{2Q_0}}\ppin{k} \frac{\Delta_{Bk}}{2\ok{}}\right.\nonumber\\&&\left.\left[i\frac{Q_0\alpha+q}{\sqrt{Q_0}}\g_{-k}(z)
+\frac{\g\p_{-k}(z)}{\sqrt{Q_0}}\left(1-\frac{\sigma^2\left(Q_0\alpha+q\right)^2}{Q_0}\right)
\right]
\right].\nonumber
\eea

Now we would like to play the same trick as before, fixing $\epsilon$ using (\ref{ep}) to eliminate all of the $q$ dependence in the biggest square brackets.  The trouble is that now the $q$ dependence now longer only appears in the combination $\epsilon$, there is also a factor of $\alpha=(\epsilon-q)/Q_0$ in front.  So, after replacing all terms $q+Q_0\alpha$ by $\epsilon$, we must also pull out this $\alpha$, yielding the form factor
\bea
\tilde{\cc}_{2,q}&=&\int dz e^{iqz}\cc_{2,\epsilon}(z)
\eea
whose  transform is
\beq
\cc_{2,\epsilon}(z)=\left(\frac{\partial_z+i\epsilon}{Q_0^{3/2}}  \right) \sigma^2 e^{-\frac{\sigma^2\epsilon^2}{2Q_0}}\ppin{k} \frac{\Delta_{Bk}}{2\ok{}}\left[i\epsilon\g_{-k}(z)
+{\g\p_{-k}(z)}\left(1-\frac{\sigma^2\epsilon^2}{Q_0}\right)
\right].
\eeq
At $\epsilon=0$, where momentum conservation selects the center of the wave packet
\beq
\cc_{2}(z)=\frac{ \sigma^2}{Q_0^{3/2}} \ppin{k} \frac{\Delta_{Bk}}{2\ok{}}{\g\pp_{-k}(z)}.
\eeq

What order is this correction?  The only dimensional constant in the normal modes $\g(x)$ is the meson mass $m$.  $\g_{-k}(z)$ is of order $O(m^0)$, its second derivative is therefore of order $O(m^2)$, while $\ok{}$ is of order $O(m)$.  The $\Delta_{Bk}$ is of order $O(m^{1/2})$ while the $k$ integral leads to another $O(m)$.  Recalling that the $1/Q_0^{3/2}$ is of order $O(g^3 m^{-3/2})$ one finds that this correction is of order
\beq
\cc_{2}(z)\sim O(g^3(\sigma^2 m)).
\eeq
This is smaller than $\cc_{1,-Q_0\alpha}$ by a power of $g^2$.   Our goal in this note is to compute all corrections of order $O(g)$, and so this correction will not be considered further.

\subsection{Leading Correction to the Kink Ground State}  \label{check}

Next, we consider the leading correction to the kink in its rest frame
\beq
\as_{0,1}=\frac{1}{(2\pi)^{1/4}\sqrt{\sigma}}e^{-\frac{\phi_0^2}{4\sigma^2}}e^{i\alpha\Lambda\p_1}\vac_1\hsp
{}_{0,1}\langle 0;\sigma|=\frac{1}{(2\pi)^{1/4}\sqrt{\sigma}}\ {}_1\langle 0|e^{-\frac{\phi_0^2}{4\sigma^2}}.
\eeq
Again, since $\df^\dag\phi(x)\df=\phi(x)+f(x)$, and $\phi(x)$ only creates or destroys one normal mode $\g_k$, while $f(x)$ is a scalar, we are only interested in terms in $\vac_1$ with zero or one normal modes excited.  There are no terms with zero normal modes excited, and so we are only interested in the terms with one.  These are
\bea
\vac_1&=&\frac{1}{\sqrt{Q_0}}\ppin{k}\left[\gamma_1^{01}(k)+\phi_0^2 \gamma_1^{21}(k)\right]B^\dag_k\vac_0\\
{}_1\langle 0|&=&\frac{1}{\sqrt{Q_0}}\ppin{k}{}_0\langle 0|\frac{B_{-k}}{2\ok{}}\left[\gamma_1^{01}(k)+\phi_0^2 \gamma_1^{21}(k)\right]\nonumber
\eea
where we have used the fact that $\gamma^*(-k)=\gamma(k)$, which in the case of these matrix elements follows from $\g_{-k}^*(x)=\g_k(x)$ and the forms of $\gamma$ in Eq.~(\ref{gam}).

\subsubsection{Derivation}

The corresponding leading correction to the form factor is
\bea
\tilde{\cc}_{3,q}&=&{}_0\langle 0;\sigma|\df^\dag \tp_q \df \as_{0,1}+{}_{0,1}\langle 0;\sigma|\df^\dag \tp_q \df \as_{0}\\
&=&{}_0\langle 0;\sigma|\tp_q  \as_{0,1}+{}_{0,1}\langle 0;\sigma|\tp_q \as_{0}\nonumber\\
&=&\int dx e^{iqx}\ppin{k} \g_k(x) \left[\frac{1}{2\ok{}}{}_0\langle 0;\sigma|B_{-k} \as_{0,1}
+{}_{0,1}\langle 0;\sigma|B^\dag_k \as_{0}
\right]\nonumber\\
&=&\frac{2}{\sqrt{Q_0}}\frac{1}{\sigma\sqrt{2\pi}}\int dx e^{iqx}\ppin{k} \frac{\g_{-k}(x)}{2\ok{}}\int dy e^{-\frac{y^2}{2\sigma^2}-i(Q_0\alpha)y/\sqrt{Q_0}}\nonumber\\
&&\times\left[{\gamma_1^{01}(k)}+y^2{\gamma_1^{21}(k)}
\right]\\
&=&\int dz e^{iqz} 
\left[
\frac{2}{\sqrt{Q_0}}\frac{1}{\sigma\sqrt{2\pi}}\int dy \left[\ppin{k} \frac{\g_{-k}\left(z-\frac{y}{\sqrt{Q_0}}\right)}{2\ok{}}\right]\right.\nonumber\\
&&\left.\times \left[{\gamma_1^{01}(k)}+y^2{\gamma_1^{21}(k)}
\right] e^{-\frac{y^2}{2\sigma^2}-i(Q_0\alpha+q)y/\sqrt{Q_0}}
\right].\nonumber
\eea
This time we will keep only the constant term in the power series expansion of $\g_{-k}$, as this term will not vanish even at $q=-Q_0\alpha$.  Performing the $y$ integral and fixing $\epsilon$ we find
\bea
\tilde{\cc}_{3,q}&=&\int dz e^{iqz} \cc_{3,\epsilon}(z)
\eea
whose Fourier transform, with $\epsilon$ fixed, is
\beq
\cc_{3,\epsilon}(z)=\frac{2}{\sqrt{Q_0}}e^{-\frac{\sigma^2\epsilon^2}{2Q_0}}
\ppin{k} \frac{\g_{-k}\left(z\right)}{2\ok{}} \left[{\gamma_1^{01}(k)}+\sigma^2\left(1-\frac{\sigma^2\epsilon^2}{Q_0}\right){\gamma_1^{21}(k)}
\right]. \label{c30}
\eeq


We can simplify the $k$ integrals by using the exact forms of $\gamma_1$ from Ref.~\cite{me2loop}
\beq
\gamma_1^{21}(k)=\frac{\ok{}\Delta_{kB}}{2}\hsp
\gamma_1^{01}(k)=\frac{\Delta_{kB}}{2}-\frac{g\sqrt{Q_0}}{2\ok{}}\int dx \V3 \I(x) \g_k(x) \label{gam}
\eeq
where $\I(x)$ is the loop factor \cite{mewick}
\beq
\I(x)=\pin{k}\frac{\left|{\g}_{k}(x)\right|^2-1}{2\omega_k}+\sum_S \frac{\left|{\g}_{S}(x)\right|^2}{2\omega_k}.\label{di}
\eeq
The $\gamma_1^{21}$ integral is
\bea
\ppin{k} \frac{\g_{-k}(z)}{2\ok{}}\gamma_1^{21}(k)&=&\frac{1}{4}\int dx \ppin{k} \g_{-k}(z) \g_k(x) \g\p_B(x)\\
&=&\frac{1}{4}\int dx \left(\delta(x-z)-\g_B(x) \g_B(z)  \right)\g\p_B(x)\nonumber\\
&=&\frac{\g\p_B(z)}{4}=\frac{f\pp_B(z)}{4\sqrt{Q_0}}.
\nonumber
\eea
Substituting this into (\ref{c30}) one finds that the $\gamma_1^{21}$ term is equal to minus $\cc_{1,\epsilon}(z)$ as given in Eq.~(\ref{c10}).  Therefore the sum of the two corrections is
\bea
\cc_{1,\epsilon}(z)&+&\cc_{3,\epsilon}(z)=\frac{2}{\sqrt{Q_0}}e^{-\frac{\sigma^2\epsilon^2}{2Q_0}}
\ppin{k} \frac{\g_{-k}\left(z\right)}{2\ok{}}{\gamma_1^{01}(k)}\label{c3}\\
&=&\frac{1}{\sqrt{Q_0}}e^{-\frac{\sigma^2\epsilon^2}{2Q_0}}
\int dx\ppin{k} \frac{\g_{-k}\left(z\right)\g_k(x)}{2\ok{}} \left[\g_B\p(x)-\frac{g\sqrt{Q_0}\V3\I(x)}{\ok{}}
\right].\nonumber
\eea
At $\epsilon=0$ this reduces to
\bea
\cc_{1}(z)+\cc_{3}(z)
&=&\frac{1}{\sqrt{Q_0}}\int dx\ppin{k} \frac{\g_{-k}\left(z\right)\g_k(x)}{2\ok{}} \left[\g_B\p(x)-\frac{g\sqrt{Q_0}\V3\I(x)}{\ok{}}
\right].\nonumber
\eea

\subsubsection{Interpretation}

The last term of (\ref{c3}) resembles the quantum correction to this matrix element computed in Eq.~(6.5) of Ref.~\cite{gervaisff75} and Eq.~(5.4) of Ref.~\cite{gj75} in the case of the $\phi^4$ model.   It is not quite the same, their result corresponds to ours without the $-1$ in the numerator of Eq.~(\ref{di}).  This $-1$ is necessary for $\I(x)$ to be finite, and in our calculation it results from the plane wave normal ordering in our defining Hamiltonian.  In Ref.~\cite{gervaisff75} instead of normal ordering, the authors use a mass counterterm, which they explain needs to be added to their result.  The addition of this term is straightforward using the Feynman rules that they provide, and we have checked that it indeed yields the $-1$ and so, with its inclusion, our results agree.


The map between our notation and that of Gervais, Jevicki and Sakita in Ref.~\cite{gervaisff75} is as follows, with our notation on the right hand side of each equation
\bea
&&f_{\rm{GJS}}(z)=\cc_{1}(z)+\cc_{3}(z)\hsp 
\tilde{G}_{\rm{GJS}}(0;z,x)=\ppin{k}\frac{\g_{-k}(z)\g_k(x)}{\ok{}^2}\\
&&\frac{3\phi_{0,\rm{GJS}}(x)}{\lambda_{\rm{GJS}}^2}=\frac{\V3}{2}\hsp
G_{\rm{GJS}}(0;x,x)\sim \I(x)
\nonumber
\eea
where the $\sim$ symbol reminds the reader about the $-1$ that results from our normal ordering and their counterterm.   The path integral derivation, used there, is quite straightforward and robust.  Schematically, the kink Lagrangian density contains the terms \cite{mewick,me2loop}
\beq
\phi \left(\Box+\V2\right)\phi+ \frac{\V3\I}{2} \phi.
\eeq
Completing the square, one finds that the squared term is
\beq
\phi+\frac{\V3\I/2}{\Box+\V2}
\eeq
and so the expectation value of $\phi$, our form factor, is
\beq
-\frac{\V3\I/2}{\Box+\V2}. \label{c3sc}
\eeq
As a result of Eq.~(\ref{sl}), the inverse of $(\Box+\V2)$ is just $\frac{1}{\ok{}^2}$ sandwiched between the complete set of $(-\partial_x^2+\V2)$ eigenvectors $\g_k$.  Eq.~(\ref{c3sc}) is then just the right hand side of our master formula (\ref{c3}).

This last term is also equal to the quantum correction to the classical kink solution $f$ in $\df$ which eliminates a tadpole term in $H_3$ when normal mode normal ordered, as found in Eq.~(3.17) of Ref.~\cite{megirini}.  If one instead interprets both terms as a quantum correction to $f(x)$ in $\df$, then the corresponding $\gamma_0^{01}$ would vanish.  More precisely, if $F(z)$ is the Fourier transform of the form factor, then $\mathcal{D}_F$ could be used to define a quantum-improved kink sector.  Our choice of state $\df\as$ in the defining Hilbert space is independent of this choice, as is the choice of state $\as$ in the original kink sector.  Therefore the corresponding state 
\beq
\as_F=\mathcal{D}_F^\dag \df\as
\eeq
depends on the choice of $F$.   Then, leaving implicit the projection to $q=-Q_0\alpha$ to avoid clutter,
\bea
{}_F\langle\sigma;\alpha|\phi(x)\as_F&=&-F(x)+{}_F\langle\sigma;\alpha|\left(\phi(x)+F(x)\right)\as_F\\
&=&-F(x)+{}_F\langle\sigma;\alpha|\mathcal{D}_F^\dag \phi(x)\mathcal{D}_F\as_F\nonumber\\
&=&-F(x)+\langle\sigma;\alpha|\df^\dag\phi(x)\df\as=-F(x)+F(x)=0.\nonumber
\eea
Therefore $F(x)$ is the quantum modified kink solution in the sense that the tadpole ${}_F\langle\sigma;\alpha|\phi(x)\as_F$ vanishes.


The first term of (\ref{c3}) on the other hand is subdominant by a factor of $m/\ok{}$.  The momentum smearing of our wave packet is much greater than $m$ \cite{meimpulso} and so if the values of $k$ that dominate this integral are of order the kink momentum, then this factor is small.  Therefore this term, while present for a wave packet of type $\as$, may well be a consequence of the smearing and so it may have no analogue in the $\sigma=\infty$ case.




As $\gamma_1^{01}$ is $O(m^{1/2})$, in all the order is
\beq
\cc_{1}(z)+\cc_{3}(z)\sim O(g)
\eeq
and it is suppressed with respect to the leading form factor by order $O(g^2)$.  

\subsection{Leading Correction to the Normalization}

\subsubsection{The Normalization}

So far we have fixed the normalization constant $\mathcal{N}$ to unity, as is correct at leading order.  More generally, it is fixed by the normalization condition (\ref{norm}).

As the boost is unitary, one can fix the normalization $\mathcal{N}$ by normalizing the at rest wave packets
\beq
\langle 0;\sigma|0;\sigma\rangle=1.
\eeq
The corrections to the wave packets considered above have a single normal mode excited, whereas the leading wave packet $\as_0$ has no normal modes excited.  Therefore $\as_0$ is orthogonal to these corrections.

To order $g^2$, the normalization condition is then
\beq
1=\langle 0;\sigma|0;\sigma\rangle=\mathcal{N}{}_0\langle 0;\sigma|0;\sigma\rangle_0+{}_{0,1}\langle 0;\sigma|0;\sigma\rangle_{0,1}=\mathcal{N}+{}_{0,1}\langle 0;\sigma|0;\sigma\rangle_{0,1}.
\eeq

\subsubsection{Calculating the Normalization}

This can be evaluated to yield
\bea
\mathcal{N}-1&=&-{}_{0,1}\langle 0;\sigma|0;\sigma\rangle_{0,1}=-\frac{1}{\sigma\sqrt{2\pi}}{}_1\langle 0|e^{-\frac{\phi_0^2}{2\sigma^2}}\vac_1\\
&=&-\frac{1}{Q_0 \sigma\sqrt{2\pi}}\ppink{2}\nonumber\\
&&\times{}_0\langle 0|\frac{B_{-k_1}}{2\ok{1}}\left[\gamma_1^{01}(k_1)+\phi_0^2 \gamma_1^{21}(k_1)\right]
\left[\gamma_1^{01}(k_2)+\phi_0^2 \gamma_1^{21}(k_2)\right]B^\dag_{k_2}e^{-\frac{\phi_0^2}{2\sigma^2}}\vac_0\nonumber\\
&=&-\frac{1}{Q_0 \sigma\sqrt{2\pi}}\int dy e^{-\frac{y^2}{2\sigma^2}} \ppin{k}\frac{1}{2\ok{}}\left[\gamma_1^{01}(-k)+y^2 \gamma_1^{21}(-k)\right]
\left[\gamma_1^{01}(k)+y^2 \gamma_1^{21}(k)\right].\nonumber
\eea
Defining the two by two matrix
\beq
M_{ij}=\ppin{k}\frac{\gamma_1^{i1}(-k)\gamma_1^{j1}(k)}{2\ok{}} \label{m}
\eeq
this simplifies to
\beq
\mathcal{N}-1=\frac{M_{00}+2\sigma^2 M_{02}+3\sigma^4 M_{22}}{Q_0}.
\eeq
This is of order $O(g^2)$, and so the correction to the form factor due to normalization is of the same order in the perturbative expansion in $g$ as the other corrections considered above.

The corresponding correction to the form factor is
\beq
\cc_{4,\epsilon}(z)=\left(\mathcal{N}-1\right)\ff_{0,\epsilon}(z)=\frac{f(z)}{Q_0}\left(M_{00}+2\sigma^2 M_{02}+3\sigma^4 M_{22}\right)e^{-\frac{\sigma^2\epsilon^2}{2Q_0}}. \label{c4}
\eeq
As $M_{ij}$ is of order $O\left(m^{(i+j)/2})\right)$, these three terms are of order $O(g),\ O(g(\sigma^2 m))$ and $O(g(\sigma^2 m)^2)$ respectively.  The first term is therefore of the same order as $\cc_3$, while the others dominate if $\sigma^2 m>>1$.

Above we have computed the corrections to the normalization resulting from the leading corrections to the ground state with a single excited normal mode.  There is also \cite{me2loop} a correction with two excitations and one $\phi_0$ and one with three excitations and no $\phi_0$, which are mutually orthogonal and orthogonal to the correction above.  These corrections are identical to those computed above with $M$ in Eq.~(\ref{m}) defined using $\gamma_1^{12}(k_1,k_2)$ and $\gamma_1^{03}(k_1,k_2,k_3)$ from Ref.~\cite{me2loop}, with a single $\sigma^2$ in the first case and no $\sigma$-dependence in the second.

\subsubsection{The Leading Correction}

There is another correction at leading order, the matrix element
\beq
\tilde{\cc}_{5,q}={}_{0,1}\langle 0;\sigma|\tilde{f}_q|\alpha;\sigma\rangle_{0,1}.
\eeq
As $\tilde{f}_q$ is a scalar, the bra and ket must have the same quantum numbers and so there is a one to one correspondence between the corrections $\tilde{\cc}_ 5$ and the leading corrections to $\mathcal{N}-1$ computed above.

For concreteness, let us consider the contributions to the states with a single normal mode.  The generalization to the other components is trivial.  At leading order, only the term $\Lambda\p_1$ contributes to the boost operator and so our approximation is
\bea
\tilde{\cc}_{5,q}&=&\frac{1}{\sigma\sqrt{2\pi}}{}_1\langle 0|\tilde{f}_q e^{-\frac{\phi_0^2}{2\sigma^2}-i\alpha\sqrt{Q_0}\phi_0}\vac_1\\
&=&\int dx \frac{f(x)e^{iqx}}{Q_0 \sigma\sqrt{2\pi}}\int dy e^{-\frac{y^2}{2\sigma^2}-i\alpha\sqrt{Q_0}y} \ppin{k}\frac{1}{2\ok{}}\left[\gamma_1^{01}(-k)+y^2 \gamma_1^{21}(-k)\right]
\left[\gamma_1^{01}(k)+y^2 \gamma_1^{21}(k)\right]\nonumber\\
&=&\int dz \frac{e^{iqz}}{Q_0 \sigma\sqrt{2\pi}}\int dy f\left(z-
\frac{y}{\sqrt{Q_0}}\right)e^{-\frac{y^2}{2\sigma^2}-i\epsilon y/\sqrt{Q_0}} \left[M_{00}+2y^2M_{02}+y^4M_{22}\right].
\nonumber
\eea
Again we expand $f$ about $f(z)$ and take the constant term, so $f(z-y/\sqrt{Q_0})$ is approximated by $f(z)$.  The later terms in the expansion will be subdominant in our perturbative expansion in $g$.

Thus we find
\bea
\cc_{5,\epsilon}(z)&=&\frac{f(z)}{Q_0}e^{-\frac{\sigma^2\epsilon^2}{2Q_0}}\left[M_{00}+2\sigma^2 \left(1-\frac{\sigma^2\epsilon^2}{Q_0}\right)M_{02}+3\sigma^4\left(1-2\frac{\sigma^2\epsilon^2}{Q_0}+\frac{\sigma^4\epsilon^4}{3Q_0^2}\right) M_{22}\right].
\nonumber
\eea
Adding this to the correction to $\mathcal{N}$ summarized in Eq.~(\ref{c4}), we arrive at the total normalization correction
\bea
\cc_{4,\epsilon}(z)+\cc_{5,\epsilon}(z)&=&f(z)\frac{\sigma^4\epsilon^2}{Q_0^2}e^{-\frac{\sigma^2\epsilon^2}{2Q_0}}\left[-2 M_{02}+\left(-6+\frac{\sigma^2\epsilon^2}{Q_0}\right) M_{22}\right].
\eea
In particular, we find that at $\epsilon=0$, there is no normalization correction at leading order.  It is easy to see that the same is true of contributions with two or three normal modes.  

Intuitively this cancellation is reasonable as the normalization correction arises from vacuum loops, contributing to the denominator of the matrix element, and the numerator term $\langle f\rangle$, which contributes at leading order, contains the same vacuum loops.   It is a generalization of the usual cancellation of disconnected diagrams in the numerator and denominator of a Greens function. 

\section{Delocalized Kinks} \label{delsez}

We sought to find form factors for strongly localized kinks.  However several of the terms that we found without $\sigma$-dependence agreed with results in the literature for delocalized kinks at tree level in Ref.~\cite{gj75} and even at the next order in Ref.~\cite{gervaisff75}.  This may seem strange as these terms are those which survive at $\sigma=0$ whereas delocalization is the opposite limit, $\sigma^2 m\rightarrow\infty$.  

Our explanation for this fact is as follows. Recall that $-\phi_0/\sqrt{Q_0}$ is the position operator for the kink center of mass.  Its eigenvalue $-y/\sqrt{Q_0}$ agrees with the collective coordinate, at leading order.  However, although a shift in the collective coordinate is a symmetry of the delocalized kink, at any fixed order in perturbation theory a shift in the eigenvalue $y$ of $\phi_0$ is not a symmetry of the states that we construct.  This is because our construction is perturbative in $y$.  Therefore, as $y$ grows, our solution is further from the correct solution.  In fact, when $y\sim 1/\sqrt{mg}$, corresponding to a collective coordinate of $\sqrt{g}/m$, our solution is at the radius of convergence of this expansion and so is essentially unrelated to the kink state.  As a result, to get reliable states, we fix $\sigma<<1/\sqrt{mg}$, which implies that at each $y$ in the support of our wave packet, our solution is reliable.

However, as was noted in Ref.~\cite{gervaisff75}, at each order the form factors are of the form $\int dz e^{iqz}\cc$ where $\cc$ is a function of $\alpha$ and $z$ and $z=x+y/\sqrt{Q_0}$ is the coordinate in the coordinate frame of the kink.  In particular, as delocalized kinks are momentum eigenstates, they are invariant under translations in the following sense.  One may choose a different base point, which means defining a shifted kink Hilbert space and kink operators using $\mathcal{D}_{f(x-x_0)}$ for any shift $x_0$.  The normal modes are then chosen to be those of $f(x-x_0)$.  Translation invariance now implies that the shifted kink Hamiltonian, in terms of the new normal modes, is identical to the unshifted kink Hamiltonian in terms of the old normal modes.  As a result, the kink Hamiltonian eigenstates, as functions of $\phi_0$ and $B^\dag$, are unchanged by this shift in $x$, so long as one always defines $\phi_0$ and $B^\dag$ using the normal modes corresponding to the base point considered. 

This is all true, order by order, in our approach.  However, translation invariance implies more, even nonperturbatively.  Recall that each term in the form factors is determined as an integral over $y$ of an integrand which depends on both the kink position $-y/\sqrt{Q_0}$ and the laboratory frame coordinate $x$ of the operator $\phi(x)$.  The integrand is roughly the contribution to the amplitude for the creation or annihilation of a meson at the position $x$ arising from a kink at collective coordinate\footnote{Note that the identification between $-y/\sqrt{Q_0}$ and the collective coordinate receives corrections of order $O(y^2)$, which mix terms among the integrands at various values of $y$.  Below we will reorganize the integral so that $y=0$ while $x$ varies, so that these corrections vanish.} $-y/\sqrt{Q_0}$.  Each such contribution may be written as a matrix element of $\phi(x)$ between position-eigenstate kinks, and so must be translation invariant.  In other words, the integrand  is invariant under a shift of $x$ and $y$ that preserves $z$.  


On the other hand, we found in the case of localized kinks that $\cc(z)$ is determined by an integral over $y$ such that $z=x+y/\sqrt{Q_0}$ and our perturbative expansion expressed this integral in moments of $y$.  Matching this power series in $y$ in the case of localized kinks with the $y$-independence argued above in the case of delocalized kinks, one arrives at the following conclusion.  In the case of delocalized kinks, translation invariance implies that all of the nonzero moments must vanish.  Recall that the $j$th moment gave a factor of $\sigma^j$, therefore in the delocalized case, only the $\sigma^0$ term survives.  These terms are $y$-independent and so can be calculated at $y=0$, where our perturbative expansion is reliable.  Now, to go to the delocalized limit, we need to take the limit $\sigma^2m\rightarrow\infty$, which is beyond the validity of our perturbative approach.  However the miracle is that these $\sigma^0$ terms are formally independent of $\sigma$, and so they do not change.  This leads us to identify the $\sigma=0$ terms in the form factors of localized kinks with those of delocalized kinks. 


One might object that we have included a $e^{-\phi_0^2/4\sigma^2}$ in our state, and so our state has been modified from the delocalized form.  Therefore the form factors should not agree.  This is true.  The argument above implied that it is only the terms with no $\sigma$ which need to agree.  These terms are clearly unchanged if one takes $\sigma^2 m>>1$ with $m$ fixed, in which case the kink is delocalized.  However this limit needs to be taken with care, as in Ref.~\cite{meimpulso} it was argued that our wave packets $\as$ have a momentum width much greater than the meson mass $m$.  On the other hand, the momentum eigenstates have a fixed momentum.   Therefore the $O(\sigma^0)$ terms in the localized kink form factors at an expected momentum $q$ can only be expected to agree with the delocalized form factor at a momentum smeared about $q$ with a width of at least $m$.  

This leads us to believe that the delocalized kink form factor, which naively corresponds to $\sigma^2m=\infty$, in fact is equal to our localized kink form factor at $\sigma=0$ up to corrections of order $O(m/q)$.    Physically, this means that our results for delocalized kinks will only be reliable for ultrarelativistic mesons, which have $q>>m$.  In the next section we will test this conclusion in the case of the Sine-Gordon model, where the form factor has been computed using integrability.


One might worry that this relation will break down at higher orders, where loops of virtual zero-modes will cause additional $y$ integrals.  Physically, one might think that there will be virtual processes where the kink emits some normal modes, and so its center of mass $-y/\sqrt{Q_0}$ recoils, and then it reabsorbs them.  In this case the form factor would necessarily depend on the wave function at $y\neq 0$.  While in the loop corrections that we have so far calculated we have seen many additional integrals over $k$, we have not yet seen any evidence that additional integrals over $y$ are required at any order.  Indeed, unlike integrals over $x$, integrals over $y$ do not arise from any contraction of fields that appear in the interaction terms of the kink Hamiltonian.  Virtual zero modes lead to additional powers of $\phi_0\g_B(x)$ in operators and so to $y\g_B(x)$ in matrix elements, and therefore apparently do not contribute to the form factor at $y=0$.

\section{The Sine-Gordon Model} \label{sgsez}

In this section we will provide a powerful check of our results, and on the matching suggested above to delocalized kinks.  We will compare the corrections calculated above to the exact Sine-Gordon form factor determined long ago in Ref.~\cite{weisz77} using integrability.   

\subsection{Our Result} \label{noisez}

In our notation, the Sine-Gordon model corresponds to the choice of potential
\beq
V(g\phi(x))=m^2\left(1-\cos\left(g\phi(x)\right)\right)
\eeq
which has a kink solution
\beq
f(x)=\frac{4}{g}\arctan\left(e^{mx}\right)
\eeq
with classical mass
\beq
Q_0=\frac{8m}{g^2}.
\eeq
There are no shape modes, but the zero mode and continuum modes are
\beq
g_{B}(x)=\sqrt{\frac{m}{2}}\sech\left(mx\right)\hsp
g_k(x)=\frac{e^{-ikx}{\rm{sign}}(k)}{\ok{}}\left(k-i m\tanh(m x)\right) \label{sggeq}
.  
\eeq
In Ref.~\cite{me2loop} we evaluated the combinations
\bea
\Delta_{kB}&=&\frac{i\pi\ok{}}{\sqrt{8m}}{\rm{sech}}\left(\frac{k\pi}{2m}\right){\rm{sign}}(k)\\
\int dx \V3 \I(x) \g_k(x)&=&\frac{i}{8m^2}\ \omega_k^3\sech\left(\frac{\pi k}{2m}\right){\rm{sign}}(k).\nonumber
\eea

Therefore, our leading contribution at $\epsilon=0$ is
\bea
\cc_{1}(z)+\cc_{3}(z)&=&\frac{1}{\sqrt{Q_0}}\pin{k} \frac{\g_{-k}\left(z\right)}{\ok{}}{\gamma_1^{01}(k)}\\
&=&\frac{ig}{16m}\pin{k} \frac{e^{ikz}}{\ok{}}\left(k+i m\tanh(m z)\right)\left(
{\pi}{}-\frac{\ok{}}{m}\right)\sech\left(\frac{\pi k}{2m}\right).\nonumber
\eea
Now, recall \cite{meimpulso} that the momentum smearing of our wave packet is much greater than the meson mass.  This implies that results at momentum transfer of order or less than the meson mass are likely to be dominated by the smearing, which has no analogue in the case of delocalized kinks which are momentum eigenstates.  Therefore, we can only hope for agreement with the momentum eigenstate form factor at momentum transfer $k>>m$.    Which terms dominate at $k>>m$?  Clearly $\omega_k/m$ dominates over $\pi$, and so we will approximate $(\pi-\ok{}/m)$ by $-\ok{}/m$.  However, as we will see momentarily, both terms in the $(k+im\tanh(mz))$ are equal.  One might have expected the $k$ term to dominate at large $k$, but this is not the case, as the tanh term contributes a power of $k$ when this full expression is rewritten in momentum space. 

Dropping the subdominant terms in this limit we arrive at the approximation
\bea
\cc_{1}(z)+\cc_{3}(z)&=&-\frac{ig}{16m^2}\pin{k} e^{ikz}\left(k+i m\tanh(m z)\right)\sech\left(\frac{\pi k}{2m}\right)\nonumber\\
&=&-\frac{ig}{16\pi m}\left(-i\partial_z+i m \tanh(mz)\right) \sech(mz)=-\frac{4f\pp(z)}{\pi g^2Q_0^2}. \label{noi}
\eea

\subsection{Weisz's Result}

In Ref.~\cite{weisz77}, Weisz calculated the form factor $\tilde{G}$ for $\phi\p(x)$ in momentum space, up to the overall normalization.   The overall normalization constant was computed in Ref.~\cite{smirnov92}, but we will instead simply fix the normalization constant by demanding that the leading contribution to the form factor for $\phi(x)$ is the Fourier transform of the classical solution.

In our notation, Weisz's form factor is just $-iq\tilde{\ff}_q$.  It was found to be of the form
\beq
\tilde{G}_q=\frac{\cosh(\theta/2)}{\cosh\left(\frac{\theta}{2}\left(\frac{8\pi}{g^2}-1\right)\right)}e^{\int_0^\infty dx I(x)}
\eeq
where
\beq
\frac{q}{2Q}=\pm i\cosh\left(\frac{i\pi-\theta}{2}\right)=\mp\sinh\left(\frac{\theta}{2}\right)
\eeq
and $I(x)$ will be given momentarily.  As $1/Q$ is dominated by $1/Q_0$, which is of order $O(g^2)$, the cubic correction to $\sinh$ is suppressed by $O(g^4)$, which is beyond the order that we are considering.  Thus we may approximate $\sinh$ at the linear order, yielding
\beq
\tilde{G}_q=\frac{\sqrt{1+\frac{q^2}{4Q^2}}}{\cosh\left(\frac{q}{2Q}\left(\frac{8\pi}{g^2}-1\right)\right)}e^{\int_0^\infty dx I(x)}.
\eeq
Expanding the denominator to order $O(g^2)$ we find
\beq
\frac{q}{2Q}\left(\frac{8\pi}{g^2}-1\right)\sim \frac{q}{2(Q_0+Q_1)}\left(\frac{8\pi}{g^2}-1\right)= \frac{q}{2(8m/g^2-m/\pi)}\left(\frac{8\pi}{g^2}-1\right)=\frac{q\pi}{2m}.
\eeq
The $O(g^2)$ correction vanishes because of a cancellation between $Q_0+Q_1$ and the parametrization of the Thirring coupling $8\pi/g^2-1$.  This remarkable cancellation is indeed necessary for our results to be consistent with those of Weisz.  The $q^2/(4Q^2)$ term in the numerator is already of order $O(g^4)$ and so its quantum corrections are of $O(g^6)$.  Therefore, the corrections that we are trying to match, those of order $O(g^2)$, can only arise from the $I(x)$ term.  

We will soon see that at leading order $I(x)=0$.  This implies that at leading order
\beq
\tilde{G}_q=\frac{1}{\cosh\left(\frac{q}{2Q_0}\left(\frac{8\pi}{g^2}\right)\right)}=\sech\left(\frac{q\pi}{2m}\right).
\eeq
This indeed is proportional to the Fourier transform of $g_B(x)$ in (\ref{sggeq}).  This is as expected, since $g_B(x)$ is proportional to $f\p(x)$ and this is a matrix element of $\phi\p(x)$, it is just the usual result \cite{gj75}, rederived in Sec.~\ref{classsez}, that the leading form factor is the Fourier transform of the classical solution.

The term $I(x)$ is defined to be
\beq
I(x)=\frac{1}{x}\frac{\sinh\left(\frac{x}{2}\left(1-\frac{1}{8\pi/g^2-1}\right)\right)}{\sinh\left(\frac{x}{2\left(8\pi/g^2-1\right)}\right)\cosh(x/2)}
\frac{\sin^2\left(\frac{x\theta}{2\pi}\right)}{2\sinh(x/2)\cosh(x/2)}.
\eeq
At $x>>1$ the numerator scales as $e^{x/2}$ while the denominator scales as $e^{3x/2}$ thus this drops exponentially.  As a result, the main contribution comes from $x$ of order unity or less.  As $\theta$ is small for a nonrelativistic kink, the sine term may be expanded linearly
\beq
\sin^2\left(\frac{x\theta}{2\pi}\right)\sim \left(\frac{x\theta}{2\pi}\right)^2\sim \left(\frac{x q}{2\pi Q_0}\right)^2.
\eeq
Similarly at leading order in $g$ one approximates
\beq
\sinh\left(\frac{x}{2}\left(1-\frac{1}{8\pi/g^2-1}\right)\right)\sim\sinh\left(\frac{x}{2}\right)\hsp
\sinh\left(\frac{x}{2\left(8\pi/g^2-1\right)}\right)\sim\frac{xg^2}{16\pi}.
\eeq
Assembling these approximations, we arrive at
\beq
I(x)=\frac{2q^2}{\pi g^2Q_0^2}\sech^2\left(\frac{x}{2}\right)
\eeq
and so
\beq
\int_0^\infty dx I(x)=\frac{4q^2}{\pi g^2Q_0^2}.
\eeq

How does this affect the matrix elements of $\phi(x)$?  Let us fix the normalization of $\tilde{G}_q$ by recalling that the leading order form factor is just the classical solution.  Then at leading order
\beq
\tilde{G}_q=-iq \tilde{\ff}_q= -iq\tf_q + O(g).
\eeq
Recall that the $O(g^2)$ corrections arise entirely from $I(x)$.  Then we find that up to order $O(g^2)$
\beq
\tilde{\ff}_q=\tf_q e^{\int_0^\infty dx I(x)}=\left(1+\frac{4q^2}{\pi g^2Q_0^2}\right)\tf_q.
\eeq
The Fourier transform of the $O(g^2)$ correction is obtained by replacing $q^2$ with $-\partial^2_z$
\beq
-\frac{4f\pp(z)}{\pi g^2Q_0^2}. \label{weisz}
\eeq
This agrees with the correction that we obtained in Eq.~(\ref{noi}).   Note that although the overall normalization of the form factor was ignored in this calculation, we fixed the normalization of the classical form factor to $f(z)$, which agrees with the normalization in Subsec.~\ref{noisez}.   The relative normalization between the two terms was never ignored.  Therefore, the normalization of Eq.~(\ref{weisz}) needs to agree with that of Eq.~(\ref{noi}), and indeed it does.


\section{Concluding Remarks}

We have found the leading and subleading contributions to the form factor corresponding to the emission or absorption of a meson by a kink in its ground state.  This was found in the Schrodinger picture, and so it corresponds to a matrix element at fixed time.  If the kink states were Hamiltonian eigenstates, such as $\vac$, they would be invariant, up to a phase, under time evolution and so this matrix element could also be interpreted as the amplitude for a kink in the past to evolve to a kink in the future.  However, in the delocalized case, they are not quite Hamiltonian eigenstates because of the $e^{-\phi_0^2/4\sigma^2}$ factors which localize them into wave packets.  These wave packets spread and evolve in time, and so an inclusion of time evolution in the matrix element would change the corresponding amplitude.  

In the future, we intend to use these form factors, as well as other matrix elements which can be calculated similarly, to calculated probabilities and rates for various physical processes in the one-kink sector.   While formulas such as the LSZ reduction formula for the S-matrix have not yet been established in this sector, one can nonetheless calculate arbitrary finite time probabilities using perturbation theory in the Schrodinger picture.  More precisely, one can start with an initial state $|i\rangle$ in the kink Hilbert space, act on it with $e^{-iH\p t}$ and then take its inner product with any desired final state $|f\rangle$.  This will give the amplitude for $|i\rangle$ to evolve to $|f\rangle$ in time $t$, and its norm squared is the corresponding probability.  Therefore matrix elements, of the kind considered here, can be used to calculate the phenomenology of a nonrelativistic kink together with its various excitations and any number of ultrarelativistic mesons.


In the case of exact momentum eigenstates, quantum corrections to the kink-meson scattering S-matrix have been evaluated in Ref.~\cite{uehara91,hayashi92}.  For the Sine-Gordon model these were found exactly in Ref.~\cite{zam77}.  It would be interesting to compare this with our future results on the scattering of mesons with kink wave packets.

\section* {Acknowledgement}

\noindent
JE is supported by the CAS Key Research Program of Frontier Sciences grant QYZDY-SSW-SLH006 and the NSFC MianShang grants 11875296 and 11675223.   JE also thanks the Recruitment Program of High-end Foreign Experts for support.

\end{document}